\documentclass[12pt]{iopart}

\usepackage{graphicx}
\usepackage{epstopdf}
\usepackage{color}
\usepackage[normalem]{ulem}

\newcommand{\bra}[1]{\left< #1 \right\vert}
\newcommand{\ket}[1]{\left\vert #1 \right>}
\newcommand{\pare}[1]{\left( #1 \right)}
\newcommand{\abs}[1]{\left\vert #1 \right\vert}
\newcommand{\cor}[1]{\left[ #1 \right]}

\newcommand{\ave}[1]{\left\langle #1 \right\rangle}

\begin{document}


\title[Two-particle quantum correlations in stochastically-coupled networks]{Two-particle quantum correlations in stochastically-coupled networks}

\author{Roberto de J. Le\'on-Montiel$^{1,*}$ \footnote{These authors contributed equally to this work},Vicen\c{c} M\'{e}ndez$^2$ $\ddagger$, Mario~A.~Quiroz-Ju\'arez$^1$, Adrian Ortega$^3$, Luis Benet$^4$, Armando~Perez-Leija$^{5,6}$ and Kurt Busch$^{5,6}$}

\address{$^1$Instituto de Ciencias Nucleares, Universidad Nacional Aut\'onoma de M\'exico,\\ Apartado Postal 70-543, 04510 Cd. Mx., M\'exico}
\address{$^2$Grup de F\'{i}sica Estad\'{i}stica, Departament de F\'{i}sica, Universitat Aut\`{o}noma de Barcelona, 08193 Barcelona, Spain}
\address{$^3$Departamento de F\'{i}sica, Universidad de Guadalajara, Blvd. Marcelino Garc\'{i}a Barragan y Calzada Ol\'{i}mpica, C.P. 44840, Guadalajara, Jalisco, M\'{e}xico.}
\address{$^4$Instituto de Ciencias F\'isicas,
Universidad Nacional Aut\'onoma de M\'exico, Apartado Postal 48-3, 62251 Cuernavaca, Morelos, M\'exico}
\address{$^5$Max-Born-Institut, Max-Born-Stra{\ss}e 2A, 12489 Berlin, Germany}
\address{$^6$Humboldt-Universit\"{a}t zu Berlin, Institut f\"{u}r Physik, AG Theoretische Optik Photonik, Newtonstra{\ss}e 15, 12489 Berlin, Germany}

\ead{roberto.leon@nucleares.unam.mx$^*$}
\vspace{10pt}

\begin{abstract}
Quantum walks in dynamically-disordered networks have become an invaluable tool for understanding the physics of open quantum systems.
In this work, we introduce a novel approach to describe the dynamics of indistinguishable particles in noisy quantum networks. By making use of stochastic calculus, we derive a master equation for the propagation of two non-interacting correlated particles in tight-binding networks affected by off-diagonal dynamical disorder. We show that the presence of noise in the couplings of a quantum network creates a pure-dephasing-like process that destroys all coherences in the single-particle Hilbert subspace. Remarkably, we find that when two or more correlated particles propagate in the network, coherences accounting for particle indistinguishability are robust against the impact of noise, thus showing that it is possible, in principle, to find specific conditions for which many indistinguishable particles can traverse dynamically-disordered systems without losing their ability to interfere. These results shed light on the role of particle indistinguishability in the preservation of quantum coherence in dynamically-disordered quantum networks.

\end{abstract}
\maketitle
%
%
%
%
%

\section{Introduction}

The study of quantum random walks in noisy environments have played a fundamental role in understanding non-trivial quantum phenomena observed in an interdisciplinary framework of studies ranging from biology \cite{rebentrost_dephtransp,chin2010}, chemistry \cite{park2016}, and electronics \cite{roberto2015}, to photonics \cite{saikin2013,viciani2015,biggerstaff2016,caruso2016} and ultracold matter \cite{schonleber2015,trautmann2017}. For many years, most of the research efforts had been focused on the propagation of single particles; however, a great interest in describing the dynamics of correlated particles in noisy systems has recently arisen \cite{rigovacca2016,siloi2017,rossi2017}, mainly because it has been recognized that many-particle quantum correlations can be preserved in noisy networks by properly controlling the initial state of the particles, their statistics, indistinguishability or their type of interaction \cite{bromley2015,beggi2016}.

In general, the interesting features in the dynamics of quantum correlated particles traversing noisy networks are due to the tunneling amplitudes in the associated Hamiltonians. Therefore, including noise into the off-diagonal elements of the Hamiltonian allows one to assess the effects of decoherence and noise.
On many occasions, when describing the evolution of correlated particles in network systems affected by non-dissipative noise, a physically accurate result can be obtained after averaging over many realizations of the noisy walks. In other words, in most cases, one does not have a master equation to analytically describe the phenomenon under study. Indeed, this represents a serious problem, specially in cases where the number of particles or network sites is extremely large. In such scenarios, computing the evolution of the system quickly becomes a computationally demanding task, which can only be tackled by developing sophisticated computer algorithms \cite{piccinini2017}. Consequently, most of the work is generally focused on optimizing numerical approaches, and the physical interpretation of the noise effects are sometimes overlooked.

In the present work we introduce a novel approach to study quantum walks in noisy systems. We use stochastic calculus to derive a master equation for the propagation of two correlated particles in a quantum network affected by off-diagonal dynamical disorder. By using our results, we show that off-diagonal noise produces an effective pure-dephasing-like process that destroys all coherences in a single-particle quantum walk. Remarkably, we find that when two or more indistinguishable particles propagate in a noisy system, coherences accounting for particle indistinguishability are robust against the dephasing-like process. These results elucidate the role of particle indistinguishability in the preservation of quantum coherence in systems that interact with a noisy environment.

\section{Single-Particle Dynamics}
We start by describing the dynamics of a single particle in a quantum network affected by random fluctuations in the coupling between sites. In this situation, the time evolution of the single-particle wavefunction at the $n$th site, $\psi_{n}$, is given by the stochastic Schr\"{o}dinger equation (with $\hbar = 1$)

\begin{equation}\label{Eq:tight-binding}
\frac{d\psi_{n}}{dt} = -i\omega_{n}\psi_{n} - i\sum_{m\neq n}\kappa_{nm}\pare{t}\psi_{m},
\end{equation}
where $\omega_{n}$ stands for the energy of the $n$th site, and the coupling between them is given by $\kappa_{nm}\pare{t} = \kappa_{nm} + \phi_{nm}\pare{t}$, with $\phi_{nm}\pare{t}=\phi_{mn}\pare{t}$ describing a white-noise process with zero average, that is, $\ave{\phi_{nm}\pare{t}} = 0$, and $\ave{\phi_{nm}\pare{t}\phi_{jl}\pare{t'}} = \gamma_{nm}\delta_{nm,jl}\delta\pare{t-t'}$. Here $\delta_{nm,jl} = \delta_{nj}\delta_{ml} + \delta_{nl}\delta_{mj}$, with $\delta_{nm}$ being the Kronecker delta. $\gamma_{nm}$ denotes the noise intensity, that is, how strong the stochastic fluctuations are, and $\ave{\cdots}$ denotes averaging over the noise realizations.

Following a treatment equivalent to the one used in  Refs. \cite{eisfeldII_dephtransp, roberto_fmo}, where fluctuations are introduced in the site-energies rather than the couplings, we can obtain a master equation for a stochastically-coupled network by taking the time derivative of $\rho_{nm}\pare{t} =\ave{\psi_{n}\psi_{m}^{*}}$. Thus, by using Eq. (\ref{Eq:tight-binding}), we can write

\begin{eqnarray}\label{Eq:master1}
\frac{d\rho_{nm}}{dt} &=& \ave{\psi_{n}\frac{d\psi_{m}^{*}}{dt} + \psi_{m}^{*}\frac{d\psi_{n}}{dt}}, \nonumber \\
& = & -i\pare{\omega_{n}-\omega_{m}}\rho_{nm} + i\sum_{j}\kappa_{mj}\rho_{nj} - i\sum_{j}\kappa_{nj}\rho_{jm} \nonumber \\
& & -i\sum_{j}\sqrt{\gamma_{mj}}\ave{\psi_{n}\psi_{j}^{*}\eta_{mj}\pare{t}} + i\sum_{j}\sqrt{\gamma_{nj}}\ave{\psi_{j}\psi_{m}^{*}\eta_{nj}\pare{t}},
\end{eqnarray}

where we have defined a new stochastic variable $\eta_{nm}\pare{t} = -\phi_{nm}\pare{t}/\sqrt{\gamma_{nm}}$, which satisfies the conditions $\ave{\eta_{nm}\pare{t}} = 0$, and $\ave{\eta_{nm}\pare{t}\eta_{jl}\pare{t'}} = \delta_{nm,jl}\delta\pare{t-t'}$. Notice that Eq. (\ref{Eq:master1}) is not yet complete, as it remains to compute the correlation functions of the last two terms. To do so, we employ the Novikov's theorem \cite{novikov1965,vicenc2011}, which for the fourth term on the right hand side of Eq. (\ref{Eq:master1}) takes the form
\begin{eqnarray}
\ave{\psi_{n}\psi_{j}^{*}\eta_{mj}\pare{t}} &=& \sum_{pq}\int dt'\ave{\eta_{mj}\pare{t}\eta_{pq}\pare{t'}} \ave{\frac{\delta\cor{\psi_{n}\pare{t}\psi_{j}^{*}\pare{t}}}{\delta\eta_{pq}\pare{t'}}}, \nonumber \\
& = & \frac{1}{2}\sum_{pq}\delta_{mj,pq}\ave{\frac{\delta\cor{\psi_{n}\pare{t}\psi_{j}^{*}\pare{t}}}{\delta\eta_{pq}\pare{t}}}. \label{Eq:Novikov1}
\end{eqnarray}
Here, it is worth remarking that the operator $\delta / \delta\eta_{pq}\pare{t}$ stands for the functional derivative with respect to the stochastic process, whose solution can be obtained by noting that
\begin{eqnarray}
\psi_{n}\pare{t}\psi_{m}^{*}\pare{t} &=& \int_{0}^{t}dt'\Bigg[ f\pare{\psi_{n}\psi_{m}^{*},...} - \left. i\sum_{r}\sqrt{\gamma_{mr}}\psi_{n}\psi_{r}^{*}\eta_{mr}\pare{t} \right. \nonumber \\
 & & \hspace{14mm} + \left. i\sum_{r}\sqrt{\gamma_{nr}}\psi_{r}\psi_{m}^{*}\eta_{nr}\pare{t} \right]. \label{Eq:solution}
\end{eqnarray}
The function $f\pare{\psi_{n}\psi_{m}^{*},...}$ contains all terms that do not depend on stochastic variables. Then, by using Eq. (\ref{Eq:solution}) we obtain
\begin{eqnarray}
\frac{\delta\cor{\psi_{n}\pare{t}\psi_{j}^{*}\pare{t}}}{\delta\eta_{pq}\pare{t}} &=& -i\sum_{r}\sqrt{\gamma_{jr}}\psi_{n}\psi_{r}^{*}\delta_{jr,pq} + i\sum_{r}\sqrt{\gamma_{nr}}\psi_{r}\psi_{j}^{*}\delta_{nr,pq}, \label{Eq:derivative1}
\end{eqnarray}
where we have used of the relation $\delta\eta_{jr}/\delta\eta_{pq} = \delta_{jr,pq}$. We can now substitute Eq. (\ref{Eq:derivative1}) into Eq. (\ref{Eq:Novikov1}) to find
\begin{eqnarray}
\ave{\psi_{n}\psi_{j}^{*}\eta_{mj}\pare{t}} &=& -\frac{i}{2}\sum_{r}\sqrt{\gamma_{jr}}\rho_{nr}\delta_{jr,mj} + \frac{i}{2}\sum_{r}\sqrt{\gamma_{nr}}\rho_{rj}\delta_{nr,mj}. \label{Eq:correlation1_1}
\end{eqnarray}
Similarly, the fifth term on the right hand side of Eq. (\ref{Eq:master1}) is found to be
\begin{eqnarray}
\ave{\psi_{j}\psi_{m}^{*}\eta_{nj}\pare{t}} &=& -\frac{i}{2}\sum_{r}\sqrt{\gamma_{mr}}\rho_{jr}\delta_{mr,nj} + \frac{i}{2}\sum_{r}\sqrt{\gamma_{jr}}\rho_{rm}\delta_{jr,nj}. \label{Eq:correlation1_2}
\end{eqnarray}
Finally, by substituting  Eqs. (\ref{Eq:correlation1_1})-(\ref{Eq:correlation1_2}) into  Eq. (\ref{Eq:master1}), we obtain
\begin{eqnarray}
\frac{d\rho_{nm}}{dt} &=& -\cor{i\pare{\omega_{n} - \omega_{m}} +\frac{1}{2}\sum_{j}\pare{\gamma_{nj}+\gamma_{mj}}}\rho_{nm} \nonumber \\
& & \hspace{2mm} + i\sum_{j}\pare{\kappa_{mj}\rho_{nj} - \kappa_{nj}\rho_{jm}} + \gamma_{nm}\rho_{nm} + \delta_{nm}\sum_{j}\sqrt{\gamma_{nj}\gamma_{mj}}\rho_{jj},  \label{Eq:density_1}
\end{eqnarray}
which corresponds to a master equation for the time evolution of a single particle in a stochastically-coupled quantum network.

\begin{figure}[t!]
\begin{center}
\includegraphics[width=11cm]{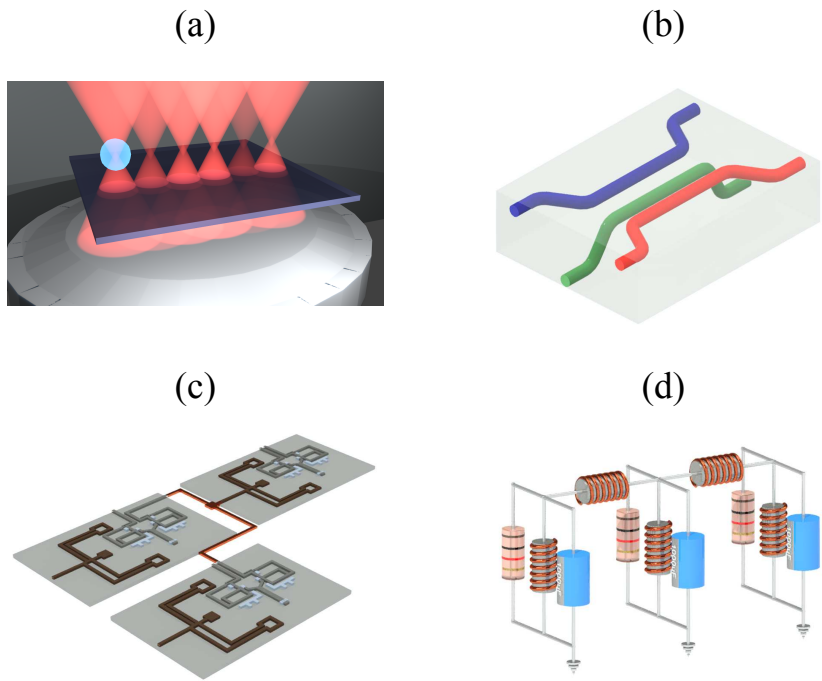}
\end{center}
\caption{Schematic representation of photonic and electronic platforms where single-excitation stochastic networks have been investigated: (a) Optical tweezers, (b) Waveguides, (c) Superconducting circuits, and (d) Electrical-circuit arrays.}
\label{Fig:network}
\end{figure}

\begin{figure}[t!]
\begin{center}
\includegraphics[width=15cm]{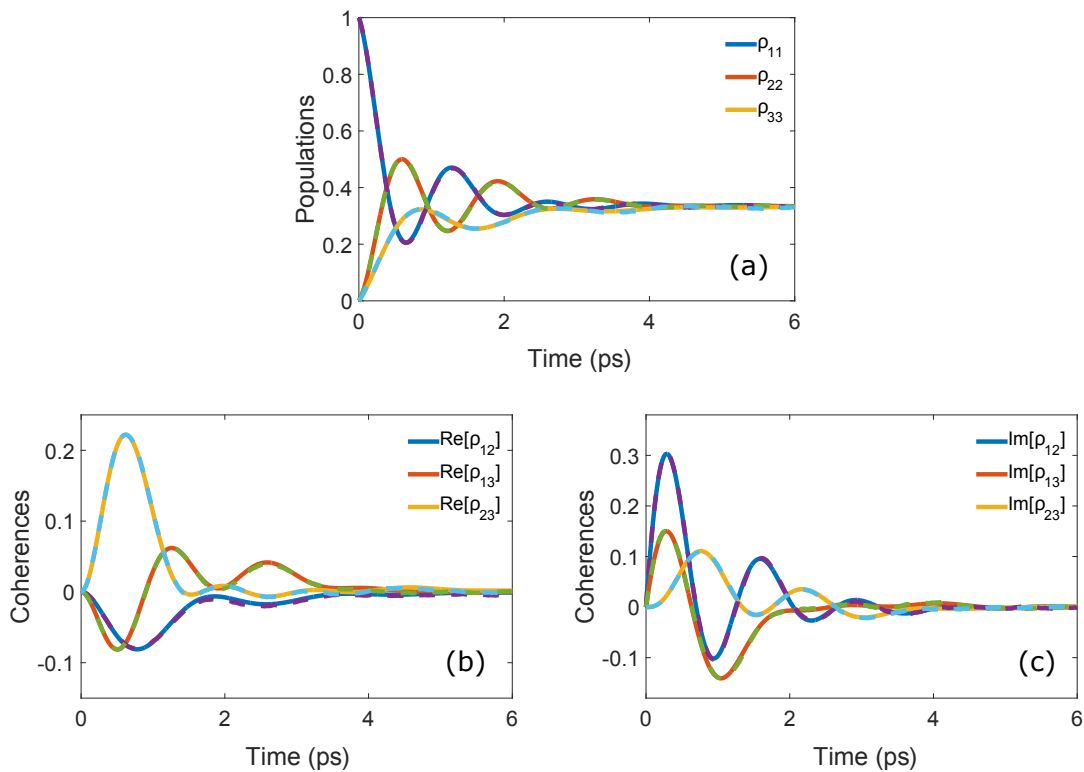}
\end{center}
\caption{Dynamics of a single excitation injected into site 1 of a stochastically-coupled three-site quantum network. (a) Time evolution of the population in each of the sites; (b) and (c) show the real and imaginary parts of the coherence (off-diagonal) terms, respectively. The solid line corresponds to the solution using the derived master equation [(\ref{Eq:density_1})]; whereas the dashed line shows the numerical solution of  (\ref{Eq:tight-binding}) obtained by averaging 10,000 realizations. In both cases, we have set the dephasing rates to $\gamma_{12}=\gamma_{13}=\gamma_{23} = 0.38 \; $ps$^{-1}$.}
\label{Fig:network2}
\end{figure}

To ellucidate the effects of the stochastic coupling between sites, we now compute the dynamics of a single excitation in a fully connected network composed by three sites with energies $\omega_{1}=\omega_{2}=\omega_{3}=5\;$ ps$^{-1}$. The couplings between them are set to $\kappa_{12}=2  \;$ ps$^{-1}$, and $\kappa_{13}=\kappa_{23}=1 \;$ ps$^{-1}$. Figure \ref{Fig:network} shows some examples of platforms where single-excitation stochastic networks have been successfully implemented, namely optical tweezers \cite{roberto2017}, waveguide arrays \cite{armando2018}, superconducting circuits \cite{Chin2018}, and electrical-circuit arrays \cite{alan2016}. The time evolution of the diagonal (populations) and off-diagonal (coherences) elements of the system's density matrix, solved by means of Eq.~(\ref{Eq:density_1}), is shown in Figure \ref{Fig:network2}. In all figures, the dephasing rate is set to $\gamma_{12}=\gamma_{13}=\gamma_{23} = 0.38 \; $ps$^{-1}$. For the sake of comparison, we have included the numerical solution (dashed lines) of Eq. (\ref{Eq:tight-binding}), which corresponds to the average of 10,000 random realizations, where the dephasing coefficient is defined by means of the relation \cite{laing_book,roberto_2014}: $\gamma_{nm} = \sigma_{nm}^{2}\Delta t$, with $\sigma_{nm}^{2}$ being the variance of the Gaussian distribution containing the values of the stochastic variable $\phi_{nm}\pare{t}$, and $\Delta t$ the correlation time. Notice that the effect of the fluctuating couplings is a pure-dephasing-like process that destroys the coherence between sites, thus leading to a steady state in which the regular hopping of the wavefunctions is no longer sustained, i.e., the system evolves into an incoherent delocalized state \cite{levi2014,roberto2015-2}.

\section{Two-Particle Wavefunction Dynamics}

We now turn our attention to the description of two-particle correlation dynamics. To this end, we use the concept of two-particle probability amplitude \cite{armando2018,armando2}, and derive the corresponding equations of motion for finite tight-binding networks comprising $N$ sites.

We start by noting that the probability amplitudes for a quantum particle, initialized at a site $n$, are governed by the equations \cite{armando2018,armando2}: $\frac{dU_{p,n}}{dt} = -i\omega_{n}U_{p,n} - i\sum_{r=1}^{N}\kappa_{pr}\pare{t}U_{r,n}$, where $U_{p,n}$ stands for the impulse response of the system, that is, the unitary probability amplitude for a single particle traveling from site $n$ to site $p$. As in the previous section, the coupling $\kappa_{pr}\pare{t}$ represents a Gaussian Markov process with zero average. We can then write, in terms of single-particle probability amplitudes, the two-particle probability amplitudes at sites $p$ and $q$ as: $\psi_{p,q}\pare{t} = \sum_{m=1,n=1}\xi_{m,n}\left[ U_{p,n}\pare{t}U_{q,m}\pare{t} \pm U_{p,m}\pare{t}U_{q,n}\pare{t} \right]$, where $\xi_{m,n}$ is the initial probability amplitude profile that fulfills the conditions $\sum_{m=1,n=1}\abs{\xi_{m,n}}^{2} = 1$. Notice that the sign $\pm$ determines whether the particles are bosons ($+$) or fermions ($-$), respectively. Then, by taking the time derivative of the two-particle wavefunction, we obtain the equation
\begin{eqnarray}
\frac{d\psi_{p,q}}{dt} &=& -i\pare{\omega_{p} + \omega_{q}}\psi_{p,q} - i\sum_{r}\cor{\kappa_{pr}\pare{t}\psi_{r,q} + \kappa_{qr}\pare{t}\psi_{p,r}}, \label{Eq:Psi3}
\end{eqnarray}
which describes the dynamics of two-particle quantum correlations. Notice that two-particle quantum states evolve in a Hilbert space composed by a discrete set of $N^{2}$-mode states occupied by the two particles. One important fact to highlight regarding Eq. (\ref{Eq:Psi3}) is the presence of the term $\pare{\omega_{p} + \omega_{q}}\psi_{p,q}$, which implies that during evolution the wavefunction $\psi_{p,q}$ acquires a phase that a single particle acquires when it traverses the same network twice \cite{giuseppe2013}. Indeed, such effects can be expected since we are dealing with two correlated particles \cite{klyshko1982}. Finally, we remark that the modulus squared of the two-particle wavefunction gives the probability of finding one particle at site $p$ and the other at $q$ \cite{abouraddy2001,saleh2005, abouraddy2002, bromberg2009,lebugle2015,weimann2016}.

\begin{figure}[t!]
\begin{center}
\includegraphics[width=15cm]{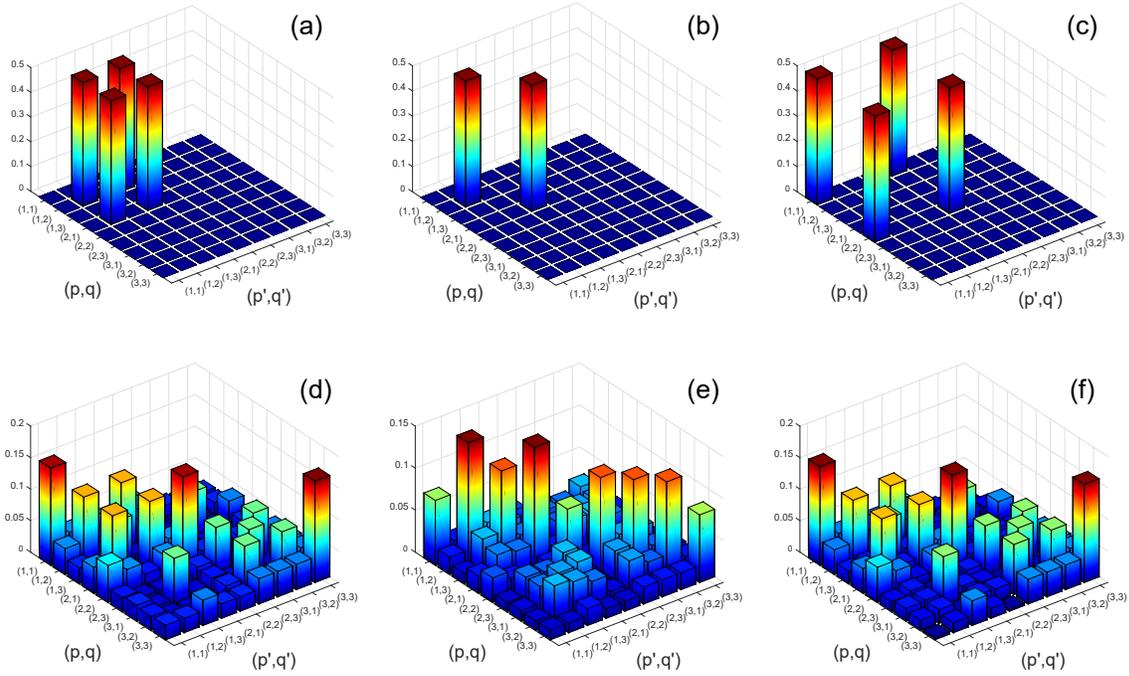}
\end{center}
\caption{Density matrices (absolute value) for (a,d) separable, (b,e) incoherent and (c,f) entangled states at $t=0$ ps and $t=1$ ps, respectively. The parameters used for the quantum network---namely site-energies, couplings and dephasing rates---are the same as in the single-particle case.}
\label{Fig:two-particle}
\end{figure}

We can now follow the same procedure as in the previous section to obtain a master equation for the two-particle wavefunction dynamics by taking the time derivative of $\rho_{pq,p'q'} = \ave{\psi_{pq}\psi_{p'q'}^{*}}$. Thus, by using (\ref{Eq:Psi3}), we obtain (see Appendix A for details)

\begin{eqnarray}
\frac{d\rho_{pq,p'q'}}{dt} &=& \Bigg[ -i\pare{\omega_{p} + \omega_{q} - \omega_{p'} - \omega_{q'}} - \gamma_{pq} - \gamma_{p'q'} \nonumber\\
& & \hspace{3mm} -\frac{1}{2}\sum_{l}\pare{\gamma_{lp} + \gamma_{lq} + \gamma_{lp'} + \gamma_{lq'}}\Bigg] \rho_{pq,p'q'} \nonumber \\
& & - i\sum_{l}\pare{\kappa_{lq}\rho_{pl,p'q'} + \kappa_{lp}\rho_{lq,p'q'}} \nonumber \\
& & + i\sum_{l}\pare{\kappa_{lq'}\rho_{pq,p'l} + \kappa_{lp'}\rho_{pq,lq'}} \nonumber \\
& & - \sum_{l}\pare{\delta_{pq}\sqrt{\gamma_{lq}\gamma_{lp}}\rho_{ll,p'q'} + \delta_{p'q'}\sqrt{\gamma_{lp'}\gamma_{lq'}}\rho_{pq,ll} } \nonumber \\
& & + \sum_{l}\pare{\delta_{qq'}\sqrt{\gamma_{lq}\gamma_{lq'}}\rho_{pl,p'l} + \delta_{qp'}\sqrt{\gamma_{lq}\gamma_{lp'}}\rho_{pl,lq'} } \nonumber \\
& & + \sum_{l}\pare{\delta_{pq'}\sqrt{\gamma_{lp}\gamma_{lq'}}\rho_{lq,p'l} + \delta_{pp'}\sqrt{\gamma_{lp}\gamma_{lp'}}\rho_{lq,lq'} } \nonumber \\
& & + \gamma_{qq'}\rho_{pq',p'q} + \gamma_{qp'}\rho_{pp',qq'} \nonumber \\
& & +\gamma_{pp'}\rho_{p'q,pq'} + \gamma_{pq'}\rho_{q'q,p'p},
\end{eqnarray}
which is the master equation that describes the time evolution of two correlated particles in a stochastically-coupled quantum network. Before considering particular examples, it is worth noting that in the following we will use the compact notation $\ket{1_{n},1_{m}}$ to represent the states where one particle is populating the site $n$ and another the site $m$, i.e. $\ket{1_{n}}\otimes\ket{1_{m}}$, whereas states $\propto\pare{ \ket{1_{n},1_{m}} + \ket{1_{m},1_{n}} }$ are symmetrized wavefunctions.

For illustrative purposes, we examine the evolution of two-particle correlations in the same network described above. As initial states we consider three different bosonic cases: (i) Two indistinguishable particles in the separable state $\ket{\psi\pare{0}}=\pare{\ket{1_{1},1_{2}} + \ket{1_{2},1_{1}}}/\sqrt{2}$, (ii) an incoherent two-distinguishable-particle state represented by $\rho\pare{0}=\pare{ \ket{1_{1},1_{2}}\bra{1_{1},1_{2}} + \ket{1_{2},1_{1}}\bra{1_{2},1_{1}} }/2$, and (iii) two particles in an entangled state $\ket{\psi\pare{0}} = \pare{\ket{1_{1},1_{1}} + \ket{1_{2},1_{2}}}/\sqrt{2}$. Figure 3 shows the evolution of the initial states at $t=1$ ps. Notice that the stochastic fluctuations affect the system in such a way that, when indistinguishable particles [Figures 3(a,d) and 3(c,f)] are injected in the system, the probability of finding both particles in the same site is the largest, that is, the photons bunch in all sites with the same probability. This effect could be thought of as a generalized Hong-Ou-Mandel effect produced by the pure-dephasing-like process. In striking contrast, when distinguishable photons are injected in the system [Figure 3(b,e)], the probability of finding them in different sites becomes larger, thus leading to an anti-bunching effect.

An important aspect to point out regarding the propagation of correlated particles in noisy quantum systems is that, recently, it has been shown that coherences arising from particle indistinguishability are robust against noise \cite{armando2018,armando2}. By making use of our model, we have verified that in the steady-state, coherences accounting for particle indistinguishability do survive the impact of stochastic fluctuations in the coupling between sites (see Appendix B for details). These results imply that it is possible, in principle, to find specific conditions for which many indistinguishable particles can traverse noisy systems without losing their ability to interfere.

Finally, notice that the generalization of our results to $N$ correlated particles is straightforward following similar steps as above by introducing the $N$-particle probability amplitude
\begin{equation}
\Psi_{p,q,r,...}\pare{t} = \sum_{a,b,c,...}^{N} \varphi_{a,b,c,...}\cor{ \chi_{a,b,c,...}^{p,q,r,...} + \chi_{a,b,c,...}^{\mathrm{per}} + ...},
\end{equation}
with $\chi_{a,b,c,...}^{p,q,r,...}=U_{p,a}\pare{t}U_{q,b}\pare{t}U_{r,c}\pare{t}...$, where $U_{m,n}$ represents the probability amplitude for each particle at site $n$ when it is injected into channel $m$. The superscript ``per'' stands for the cyclic permutations of the subscripts $p,q,r,...$ in the corresponding transition amplitudes.

\section{Conclusions}
In this work, we have derived a master equation for the propagation of correlated particles in quantum networks affected by off-diagonal dynamical disorder. Unlike commonly-used computational methods, where many stochastic trajectories are needed, our equation allows one to find the average trajectory of correlated particles in a single calculation. By using our results, we showed that the effect of introducing noise in the couplings of a quantum network leads to a dephasing-like process that destroy all coherences in the single-particle Hilbert subspace. Interestingly, we found that when two or more correlated particles propagate in a disordered network, coherences accounting for the indistinguishability of the particles endure the impact of noise. These results may help elucidating the role of particle indistinguishability to preserve quantum coherence and entanglement propagating through complex dynamically-disordered systems.

\section{Acknowledgments}
This work was supported by DGAPA-UNAM under the project UNAM-PAPIIT IA100718, and by CONACYT under the project CB-2016-01/284372.
Armando Perez-Leija and Kurt Busch acknowledge financial support by the Deutsche Forschungsgemeinschaft (PE 2602/2-2  and  BU 1107/12-2).

\appendix
\section*{Appendix}
\section*{Dynamics of many-particle quantum correlations in stochastically-coupled tight-binding networks}

In this appendix, we (i) show how to obtain the master equation describing the propagation of two correlated particles in a quantum network affected by dynamic disorder introduced in the coupling between sites, and (ii) present a quantitative comparison between our derived equation and the results obtained from the direct numerical simulation of the propagation dynamics of two correlated particles in a stochastically-coupled system.

\section{Derivation of the two-particle master equation}

We start by writing the expression for the probability amplitude dynamics of a quantum particle initiated at site $n$
\begin{equation}
\frac{dU_{q,n}}{dt} = -i\omega_{q}U_{q,n} - i\sum_{r}\kappa_{rq}\pare{t}U_{r,n},
\end{equation}
where $\omega_{n}$ stands for the energy of the $n$th site, and the coupling between the $r$th and $q$th sites is given by $\kappa_{rq}\pare{t} = \kappa_{rq} + \phi_{rq}\pare{t}$, with $\phi_{rq}\pare{t}=\phi_{qr}\pare{t}$ describing a Gaussian Markov process with zero average, that is,
\begin{equation}\label{Eq:average_stochastic}
\ave{\phi_{rq}\pare{t}} = 0,
\end{equation}
\begin{equation}\label{Eq:correlation_function}
\ave{\phi_{rq}\pare{t}\phi_{jl}\pare{t'}} = \gamma_{rq}\delta_{rq,jl}\delta\pare{t-t'}.
\end{equation}
Here $\delta_{rq,jl} = \delta_{rj}\delta_{ql} + \delta_{rl}\delta_{qj}$, with $\delta_{rq}$ being the Kronecker delta. $\gamma_{rq}$ denotes the noise intensity, that is, how strong the stochastic fluctuations are, and $\ave{\cdots}$ denotes stochastic averaging. By defining the stochastic variable $\phi_{rq}\pare{t} = -\sqrt{\gamma_{rq}}\xi_{rq}\pare{t}$, we can write
\begin{equation}
\frac{dU_{q,n}}{dt} = -i\omega_{q}U_{q,n} - i\sum_{r}\kappa_{rq}U_{r,n} + i\sum_{r}\sqrt{\gamma_{rq}}\xi_{rq}\pare{t}U_{r,n},
\end{equation}
with the properties of the stochastic variable $\xi_{rq}$ given by
\begin{equation}
\ave{\xi_{rq}\pare{t}} = 0,
\end{equation}
\begin{equation}
\ave{\xi_{rq}\pare{t}\xi_{jl}\pare{t'}} = \delta_{rq,jl}\delta\pare{t-t'}.
\end{equation}
Notice that because noise (dynamic disorder) is introduced in the couplings, we must keep in mind that $r\neq q$ and, consequently, $j\neq l$.

Now, to compute the evolution of the two-particle density matrix $\rho_{pq,p'q'}=\ave{\psi_{pq}\psi_{p'q'}^{*}}$, with $\psi_{p,q}\pare{t} = \sum_{m=1,n=1}\xi_{m,n}\left[ U_{p,n}\pare{t}U_{q,m}\pare{t}\pm U_{p,m}\pare{t}U_{q,n}\pare{t} \right]$, we first write
\begin{eqnarray}\label{Eq:rho1}
\frac{d\pare{\psi_{pq}\psi_{p'q'}^{*}}}{dt}  = & -i  \cor{\omega_{p}+\omega_{q}-\omega_{p'}-\omega_{q'}}\psi_{pq}\psi_{p'q'}^{*} \nonumber \\
& - i\sum_{l}\kappa_{lq}\psi_{pl}\psi_{p'q'}^{*} - i\sum_{l}\kappa_{lp}\psi_{lq}\psi_{p'q'}^{*} \nonumber \\
& + i\sum_{l}\kappa_{lq'}\psi_{pq}\psi_{p'l}^{*}  + i\sum_{l}\kappa_{lp'}\psi_{pq}\psi_{lq'}^{*} \nonumber \\
&- i\sum_{l}\sqrt{\gamma_{lq}}\psi_{pl}\psi_{p'q'}^{*}\xi_{lq}\pare{t} - i\sum_{l}\sqrt{\gamma_{lp}}\psi_{lq}\psi_{p'q'}^{*}\xi_{lp}\pare{t} \nonumber \\
& + i\sum_{l}\sqrt{\gamma_{lq'}}\psi_{pq}\psi_{p'l}^{*}\xi_{lq'}\pare{t} + i\sum_{l}\sqrt{\gamma_{lp'}}\psi_{pq}\psi_{lq'}^{*}\xi_{lp'}\pare{t}.
\end{eqnarray}
We can formally integrate Eq. (\ref{Eq:rho1}), and obtain
\begin{eqnarray}\label{Eq:rho2}
\psi_{pq}\psi_{p'q'}^{*} = & \int_{0}^{t}dt'\Bigg\{ f\pare{\psi_{pq}\psi_{p'q'}^{*},...}\nonumber \\
& - i\sum_{l}\sqrt{\gamma_{lq}}\psi_{pl}\pare{t'}\psi_{p'q'}^{*}\pare{t'}\xi_{lq}\pare{t'}\nonumber \\
& - i\sum_{l}\sqrt{\gamma_{lp}}\psi_{lq}\pare{t'}\psi_{p'q'}^{*}\pare{t'}\xi_{lp}\pare{t'} \nonumber \\
& + i\sum_{l}\sqrt{\gamma_{lq'}}\psi_{pq}\pare{t'}\psi_{p'l}^{*}\pare{t}\xi_{lq'}\pare{t'} \nonumber \\
&+ i\sum_{l}\sqrt{\gamma_{lp'}}\psi_{pq}\pare{t'}\psi_{lq'}^{*}\pare{t'}\xi_{lp'}\pare{t'}\Bigg\},
\end{eqnarray}
where $f\pare{\cdots}$ is a function that contains all terms that do not depend on the stochastic variables. Concurrently, we can write the average of Eq. (\ref{Eq:rho1}) as
\begin{eqnarray}\label{Eq:rho_ave}
\frac{d\ave{\psi_{pq}\psi_{p'q'}^{*}}}{dt} = -i & \cor{\omega_{p}+\omega_{q}-\omega_{p'}-\omega_{q'}}\ave{\psi_{pq}\psi_{p'q'}^{*}} \nonumber \\
&- i\sum_{l}\kappa_{lq}\ave{\psi_{pl}\psi_{p'q'}^{*}} - i\sum_{l}\kappa_{lp}\ave{\psi_{lq}\psi_{p'q'}^{*}} \nonumber \\
&+ i\sum_{l}\kappa_{lq'}\ave{\psi_{pq}\psi_{p'l}^{*}} + i\sum_{l}\kappa_{lp'}\ave{\psi_{pq}\psi_{lq'}^{*}} \nonumber \\
& - i\sum_{l}\sqrt{\gamma_{lq}}\ave{\psi_{pl}\psi_{p'q'}^{*}\xi_{lq}\pare{t}} \nonumber \\
&- i\sum_{l}\sqrt{\gamma_{lp}}\ave{\psi_{lq}\psi_{p'q'}^{*}\xi_{lp}\pare{t}} \nonumber \\
& + i\sum_{l}\sqrt{\gamma_{lq'}}\ave{\psi_{pq}\psi_{p'l}^{*}\xi_{lq'}\pare{t}} \nonumber \\
&+ i\sum_{l}\sqrt{\gamma_{lp'}}\ave{\psi_{pq}\psi_{lq'}^{*}\xi_{lp'}\pare{t}}.
\end{eqnarray}
It is clear that in order to obtain the master equation for $\rho_{pq,p'q'}\pare{t}$, we must evaluate the correlation functions in the last four terms of Eq. (\ref{Eq:rho_ave}). To do so, we invoke the Novikov's theorem \cite{novikov1965,vicenc2011}, which for the first correlation function in Eq. (\ref{Eq:rho_ave}) takes the form
\begin{eqnarray}\label{Eq:novikov1}
\ave{\psi_{pl}\psi_{p'q'}^{*}\xi_{lq}\pare{t}} &=& \sum_{rs}\int dt' \ave{\xi_{lq}\pare{t}\xi_{rs}\pare{t'}}\ave{\frac{\delta\cor{\psi_{pl}\pare{t}\psi_{p'q'}^{*}\pare{t}}}{\delta\xi_{rs}\pare{t'}}}, \nonumber \\
&=& \sum_{rs}\int dt' \delta_{lq,rs}\delta\pare{t-t'}\ave{\frac{\delta\cor{\psi_{pl}\pare{t}\psi_{p'q'}^{*}\pare{t}}}{\delta\xi_{rs}\pare{t'}}}, \nonumber \\
&=& \frac{1}{2}\sum_{rs}\delta_{lq,rs}\ave{\frac{\delta\cor{\psi_{pl}\pare{t}\psi_{p'q'}^{*}\pare{t}}}{\delta\xi_{rs}\pare{t}}}.
\end{eqnarray}
Here, we have taken into account the fact that, in the Stratonovich interpretation \cite{kampen1981}, $\int\delta\pare{t} = 1/2$. We can then use Eq. (\ref{Eq:rho2}) to write the functional derivative as
\begin{eqnarray}
\frac{\delta\cor{\psi_{pl}\pare{t}\psi_{p'q'}^{*}\pare{t}}}{\delta\xi_{rs}\pare{t}} = -i &  \sum_{\sigma}\sqrt{\gamma_{\sigma l}}\psi_{p\sigma}\pare{t}\psi_{p'q'}^{*}\pare{t}\delta_{\sigma l,rs}\nonumber \\
& - i\sum_{\sigma}\sqrt{\gamma_{\sigma p}}\psi_{\sigma l}\pare{t}\psi_{p'q'}^{*}\pare{t}\delta_{\sigma p,rs} \nonumber \\
& + i\sum_{\sigma}\sqrt{\gamma_{\sigma q'}}\psi_{pl}\pare{t}\psi_{p'\sigma}^{*}\pare{t}\delta_{\sigma q',rs}\nonumber \\
& + i\sum_{\sigma}\sqrt{\gamma_{\sigma p'}}\psi_{pl}\pare{t}\psi_{\sigma q'}^{*}\pare{t}\delta_{\sigma p',rs},
\end{eqnarray}
where we used the relation $\delta\xi_{\sigma l}/\delta\xi_{rs} = \delta_{\sigma l,rs}$. By substituting this result into Eq. (\ref{Eq:novikov1}), we can write
\begin{eqnarray}
\ave{\psi_{pl}\psi_{p'q'}^{*}\xi_{lq}\pare{t}} = - & \frac{i}{2}\sum_{\sigma}\delta_{\sigma l,lq}\sqrt{\gamma_{\sigma l}}\rho_{p\sigma , p'q'}\nonumber \\
& - \frac{i}{2}\sum_{\sigma}\delta_{\sigma p,lq}\sqrt{\gamma_{\sigma p}}\rho_{\sigma l, p'q'} \nonumber \\
& + \frac{i}{2}\sum_{\sigma}\delta_{\sigma q',lq}\sqrt{\gamma_{\sigma q'}}\rho_{pl , p'\sigma}\nonumber \\
& + \frac{i}{2}\sum_{\sigma}\delta_{\sigma p',lq}\sqrt{\gamma_{\sigma p'}}\rho_{pl , \sigma q'}.
\end{eqnarray}
Similarly, the remaining correlation functions are given by
\begin{eqnarray}
\ave{\psi_{lq}\psi_{p'q'}^{*}\xi_{lp}\pare{t}} = - & \frac{i}{2}\sum_{\sigma}\delta_{\sigma q,lp}\sqrt{\gamma_{\sigma q}}\rho_{l\sigma , p'q'}\nonumber \\
& - \frac{i}{2}\sum_{\sigma}\delta_{\sigma l,lp}\sqrt{\gamma_{\sigma l}}\rho_{\sigma q, p'q'} \nonumber \\
& + \frac{i}{2}\sum_{\sigma}\delta_{\sigma q',lp}\sqrt{\gamma_{\sigma q'}}\rho_{lq , p'\sigma}\nonumber \\
& + \frac{i}{2}\sum_{\sigma}\delta_{\sigma p',lp}\sqrt{\gamma_{\sigma p'}}\rho_{lq,\sigma q'},
\end{eqnarray}

\begin{eqnarray}
\ave{\psi_{pq}\psi_{p'l}^{*}\xi_{lq'}\pare{t}} = - & \frac{i}{2}\sum_{\sigma}\delta_{\sigma q,lq'}\sqrt{\gamma_{\sigma q}}\rho_{p\sigma , p'l} \nonumber \\
&- \frac{i}{2}\sum_{\sigma}\delta_{\sigma p,lq'}\sqrt{\gamma_{\sigma p}}\rho_{\sigma q, p'l} \nonumber \\
& + \frac{i}{2}\sum_{\sigma}\delta_{\sigma l,lq'}\sqrt{\gamma_{\sigma l}}\rho_{pq , p'\sigma} \nonumber \\
&+ \frac{i}{2}\sum_{\sigma}\delta_{\sigma p',lq'}\sqrt{\gamma_{\sigma p'}}\rho_{pq,\sigma l},
\end{eqnarray}

\begin{eqnarray}
\ave{\psi_{pq}\psi_{lq'}^{*}\xi_{lp'}\pare{t}} = - & \frac{i}{2}\sum_{\sigma}\delta_{\sigma q,lp'}\sqrt{\gamma_{\sigma q}}\rho_{p\sigma , lq'} \nonumber \\
&- \frac{i}{2}\sum_{\sigma}\delta_{\sigma p,lp'}\sqrt{\gamma_{\sigma p}}\rho_{\sigma q, lq'} \nonumber \\
& + \frac{i}{2}\sum_{\sigma}\delta_{\sigma q',lp'}\sqrt{\gamma_{\sigma q'}}\rho_{pq , l\sigma} \nonumber \\
&+ \frac{i}{2}\sum_{\sigma}\delta_{\sigma l,lp'}\sqrt{\gamma_{\sigma l}}\rho_{pq,\sigma q'}.
\end{eqnarray}
Finally, by substituting Eqs. (A.12)-(A.15) into Eq. (A.9) we obtain
\begin{eqnarray}
\frac{d\rho_{pq,p'q'}}{dt} = & -i\pare{\omega_{p} + \omega_{q} - \omega_{p'} - \omega_{q'}} \rho_{pq,p'q'}\nonumber \\
& - \frac{1}{2}\sum_{l}\cor{\pare{\gamma_{lp} + \gamma_{lq} + \gamma_{lp'} + \gamma_{lq'}} - \gamma_{pq} - \gamma_{p'q'} }\rho_{pq,p'q'} \nonumber \\
& - i\sum_{l}\pare{ \kappa_{lq}\rho_{pl,p'q'} + \kappa_{lp}\rho_{lq,p'q'} - \kappa_{lq'}\rho_{pq,p'l} - \kappa_{lp'}\rho_{pq,lq'} } \nonumber \\
& - \sum_{l}\pare{ \delta_{pq}\sqrt{\gamma_{lq}\gamma_{lp}}\rho_{ll,p'q'} + \delta_{p'q'}\sqrt{\gamma_{lp'}\gamma_{lq'}}\rho_{pq,ll} } \nonumber \\
& + \sum_{l}\pare{ \delta_{qq'}\sqrt{\gamma_{lq}\gamma_{lq'}}\rho_{pl,p'l} + \delta_{qp'}\sqrt{\gamma_{lq}\gamma_{lp'}}\rho_{pl,lq'}} \nonumber \\
&+\sum_{l}\pare{ \delta_{pq'}\sqrt{\gamma_{lp}\gamma_{lq'}}\rho_{lq,p'l} + \delta_{pp'}\sqrt{\gamma_{lp}\gamma_{lp'}}\rho_{lq,lq'} } \nonumber \\
& + \gamma_{qq'}\rho_{pq',p'q} + \gamma_{qp'}\rho_{pp',qq'} + \gamma_{pp'}\rho_{p'q,pq'} + \gamma_{pq'}\rho_{q'q,p'p},
\end{eqnarray}
which is the result shown in Eq. (10) of the main manuscript.

\section{Comparison between master equation and the direct stochastic numerical simulation}
\begin{figure}[t!]
\begin{center}
\includegraphics[width=15.5cm]{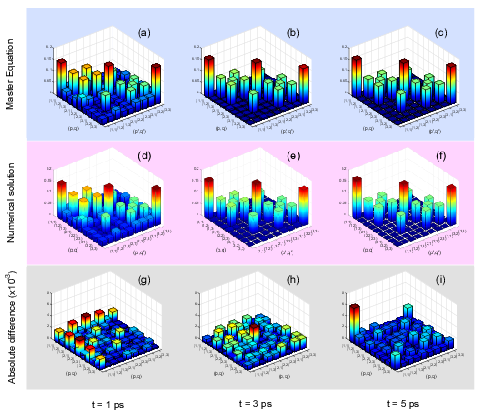}
\caption{Density matrices (absolute value) for a separable state, $\ket{\psi\pare{0}}=\pare{\ket{1_{1},1_{2}} + \ket{1_{2},1_{1}}}/\sqrt{2}$, at $t=1$ ps, $t=3$ ps, and $t=5$ ps, obtained by means of the derived master equation (a-c), and by the direct numerical evaluation of the stochastic equations (d-f). Figures B1(g-i) show the absolute difference between both solutions, $\Delta\rho = \abs{ \abs{\rho_{pq,p'q'}^{\mathrm{(master)}}} - \abs{\rho_{pq,p'q'}^{\mathrm{(numerical)}}}}$, at the corresponding evolution times.}
\end{center}
\end{figure}

We now provide a quantitative comparison between the time evolution of a two-particle state obtained by means of our derived master equation and by directly implementing the stochastic equations. Figure B1 shows the evolution of a separable state, $\ket{\psi\pare{0}}=\pare{\ket{1_{1},1_{2}} + \ket{1_{2},1_{1}}}/\sqrt{2}$, propagating in a dynamically-disordered three-site network. The parameters used for the quantum networks---namely site-energies, couplings and dephasing rates---are the same as those used for obtaining Fig. 3 of the main text. Figures B1(a-c) show the results obtained by using the derived master equation [Eq. (10) of the main text], whereas Figs. B1(d-f) show the results obtained by numerically solving Eq. (9) of the main text using the Taylor Integration package \cite{numericalpackages}. The latter were obtained by averaging over 10 000 different realizations of the two-particle random walk. It is important to highlight that the computation time required for each case was $T^{\mathrm{(master)}}_{c} = 0.521$ s, and $T^{\mathrm{(numerical)}}_{c} = 2.4$ hrs for the master equation and direct stochastic evaluation, respectively. Clearly, our derived equation improves the computation time by at least four orders of magnitude, while providing the maximum accuracy possible. For the sake of completeness, in Figs. B1(g-i), we have included the absolute difference between the absolute value of the density matrix elements obtained from the master equation and the numerical solution, i.e., $\Delta\rho = \abs{ \abs{\rho_{pq,p'q'}^{\mathrm{(master)}}} - \abs{\rho_{pq,p'q'}^{\mathrm{(numerical)}}} }$. Finally, we would like to remark that while the derived master equation provides the exact solution, the accuracy of the stochastic-computation solution strongly depends on the number of realizations being used for the average, which implies that many realizations (and therefore longer computation times) are required in order to obtain reliable numerical results. This is the reason why, when possible, one should use master equations instead of direct stochastic numerical simulations.

\section*{References}
\bibliography{References}
\bibliographystyle{iopart-num}

\end{document}